\documentclass[aps,prb,twocolumn,prabib,amsmath,amssymb,superscriptaddress]{revtex4-1}
\usepackage{graphics}
\usepackage{bm}
\usepackage{dcolumn}
\usepackage{natbib}
\usepackage{epstopdf}
\usepackage{graphicx}
\usepackage{epsfig}
\usepackage{color,xcolor}
\usepackage[pdfstartview=FitH]{hyperref}
\newcommand{\kk}{{\bf k}}

\newcommand{\qq}{{\bf q}}
\newcommand{\rr}{{\bf r}}
\begin{document}
\title{Fractional Quantum Hall States of Dipolar Gases in Chern Bands}

\author{Tian-Sheng Zeng}
\affiliation{School of Physics, Peking University, Beijing 100871, China.}

\author{Lan Yin}
\affiliation{School of Physics, Peking University, Beijing 100871, China.}
\date{\today}
\begin{abstract}
We study fermions and hardcore bosons with long range dipolar interactions at fractional fillings in a topological checkerboard lattice with short-range hoppings up to next-next-nearest neighbors \cite{Neupert2011}.   We consider the case that the dipoles are aligned in the perpendicular direction by an external field without the complication of anisotropic interaction. Using exact diagonalization, we find clear signatures of fractional quantum Hall (FQH) states at filling factors 1/3 and 1/5 for fermions (1/2 and 1/4 for bosons) in the lowest Chern band with a robust spectrum gap at moderate dipolar interaction strength. The robustness of these FQH states against long-range interaction tail and band flatness is investigated. When the dipolar interaction decreases, the fermionic FQH states turn into normal states, and the bosonic 1/4-FQH state turns into a superfluid state.  The bosonic 1/2-FQH state survives even in the absence of the dipolar interaction, but vanishes when the hard core becomes a soft core with a critical onsite repulsion.  In the thin torus limit, the static density structure factors indicates that the FQH state turns into a commensurate charge density wave (CDW) state.
\end{abstract}
\maketitle
\section{Introduction}
In a seminal paper, Haldane proposed a time-reversal breaking honeycomb lattice model as a zero magnetic field version of the integer quantum Hall effect~\cite{Haldane1988}, which is a classic example of an integer Chern insulator.  Chern insulators host topological order in the band structure with a topological invariant characterized by the Chern number~\cite{Sun2011}.   Theoretical studies suggest that the $1/3$-FQH states is the ground state at $\nu=\frac{1}{3}$ filling of the lowest Chern band with a strong nearest neighbor interaction\cite{Neupert2011,Sheng2011,Tang2011}, which has the same Hall conductance as that of the $1/3$-FQH state of two-dimensional electron gas described by the Laughlin wavefunction.  The generic wavefunctions for fractional Chern insulators can be constructed from Laughlin wave functions by one to one corrspondence between the continuous cylindrical Landau level wave function and Wannier functions~\cite{Qi2011}.  These fascinating FQH states in Chern bands have been proposed by engineering triangular optical flux lattice~\cite{Cooper2013}, square/honeycomb optical lattice using laser-induced transitions~\cite{Goldman2013}, and checkerboard lattice reduced from dipolar spin system~\cite{Yao2013} in ultra-cold $^{40}\text{K}^{87}\text{Rb}$ molecular gas, and Floquet Chern insulator~\cite{Grushin2014}. Experimentally, the Haldane honeycomb insulator has been achieved from periodically modulating system~\cite{Jotzu2014}. The integer Chern number of topological flat-bands has been revealed successfully in an all-optical artificial gauge field scheme~\cite{Aidelsburger2014}.

Numerical evidences for the $1/2$ and $1/3$-FQH states are found with large spectrum gaps induced by nearest neighbor repulsions~\cite{Neupert2011,Sheng2011,Tang2011,WBR2012,Wang2011}. It has been observed that the composite fermion state at $\nu=2/3$ is destructed upon lowering the nearest-neighbor repulsion~\cite{LiuT2013}, and fermionic $1/3$-FQH state for the honeycomb lattice model would undergo phase transitions into a Fermi liquid by tuning down the nearest-neighbor repulsion~\cite{Li2014}.  Lower filling fraction $1/4$ and $1/5$-FQH states become robust only when large next-nearest neighbor interaction is included~\cite{Sheng2011,Wang2011}, signifying the importance of long-range interaction.  The role played by long-range interaction on these FQH states is worth investigation.  In two-dimensional lattice with incommensurate magnetic flux, the spectral gap of bosonic FQH states can be enhanced by the dipole-dipole interaction~\cite{Hafezi2007}.  In the one dimensional superlattice, fractional topological states emerge as the interplay of nontrivial topological band and dipolar long-range interactions, but not for the case with only short-range interactions even in the strongly interacting limit~\cite{Xu2013}.  In a dispersionless Chern band, electrons with Coulomb repulsion exhibit a hierarchy of fractional Chern insulators~\cite{LBM2013,Venderbos2012}. Recently, it is pointed out that screened Coulomb interaction may drive fractional Chern insulator into stripe CDW order of fractional charge~\cite{Chen2014}. It is suggested that for bosons on the lowest Hofstadter subband with strong on-site interaction, weak off-site long-range dipolar interactions can enhance the robustness of non-Abelian fractional Chern insulator states~\cite{Liu2013,Bergholtz2013} with adiabatic continuity between Hofstadter and Chern insulator states~\cite{Wu2012}.

Ultracold fermionic~\cite{Chotia2012} and bosonic~\cite{Stuhler2005} dipolar gases have been realized in experiments, providing promising candidates for realizing fractional Chern insulators. Combination of both the intrinsic spin-orbit coupling of dipolar interactions and broken time-reversal symmetry would lead to a topological flat-band with Chern number two on square lattice~\cite{Peter2014}.  Yao {\it et al.} proposed to generate topological flat-band with dipolar interactions modified by the optical dressing in $^{40}\text{K}^{87}\text{Rb}$ system~\cite{Yao2013}. In a square optical lattice, antiferromagnetic and topolgocial superfluid states are proposed to exist in dipolar fermi gases~\cite{Liu2011}. The very core of this paper is to extend the previous theoretical discussion of long range interaction on FQH states~\cite{Sheng2011,Wang2011}. Here we focus our interest on effects of the long range dipolar interaction on the stability of Abelian FQH states in checkerboard lattice model, and find convincing evidence for Abelian $1/2,1/3,1/4,1/5$-FQH states of dipolar quantum gas in the lowest Chern band. In contrast, rotating quasi-2D dipolar Fermi gas with dipole-dipole interaction is predicted to exhibit $1/3$-FQH state in Landau level and crystalline order at lower fillings~\cite{Osterloh2007}.  We show that the many-body ground states at these fractional fillings are FQH states with moderate dipolar interaction. The topological properties of these states are characterized by (\textrm{i}) fractional quantized topological invariants related to Hall conductance, (\textrm{ii}) degenerate ground state manifolds under the adiabatic insertion of flux quanta, and (\textrm{iii}) the quasihole statistics. For weak interaction strength, we find quantum phase transitions from FQH states to other states, such as superfluid state of bosonic systems and Fermi liquid state of fermionic systems. We also consider the role played by aspect ratio of two dimensional torus geometry, and find that in thin torus limit, the ground state turns into a crystal phase.

This paper is organized as follows. In Sec.~\ref{chern}, we give a overview of the checkerboard lattice model with topological invariant. In Sec.~\ref{frachall}, we study dipolar gases on the checkerboard lattice with dipole direction fixed in the perpendicular direction and present numerical results of Abelian FQH states by exact diagonalization at fillings $\nu=1/2,1/3,1/4,1/5$, and discuss the properties of these ground states.  Finally, in Sec.~\ref{SD}, we summarize our results and discuss the prospect of investigating nontrivial topological states in ultracold dipolar gases.
\section{The Checkerboard Model}\label{chern}
We consider the spinless fermions or hardcore bosons on the periodic checkerboard lattice with $N_s=N_x\times N_y$ unit cells. Each unit cell consists of two inequivalent lattice sites $A$ and $B$ separated at a distance $(a/2,a/2)$ where $a=1$ is lattice constant. The Hamiltonian is given by
\begin{align}
  H_0=\sum_{\rr,\rr'}\big[&t_{1}e^{i\phi(\rr,\rr')}a_{\rr}^{\dag}b_{\rr'}
  +t_{a}(\rr,\rr')a_{\rr}^{\dag}a_{\rr'}+t_{b}(\rr,\rr')b_{\rr}^{\dag}b_{\rr'}\nonumber\\
  &+h.c.\big]+\sum_{\rr}M(a_{\rr}^{\dag}a_{\rr}-b_{\rr}^{\dag}b_{\rr}),
\end{align}
where $\rr=(x,y)$ the lattice vector, $a_{\rr}$ and $b_{\rr}$ are annihilation operators on sublattice $A$ and $B$ respectively, $M$ is an on-site staggered potential, the nearest hopping phase is given by $\phi(\rr,\rr')=\phi\times\mathrm{Sign}[(x'-x)(y'-y)]$, the next-nearest hopping parameters are given by $t_{a}(\rr,\rr')=t_2(-1)^{y'-y}$ for sublattice $A$ and $t_{b}(\rr,\rr')=t_2(-1)^{x'-x}$ for sublattice $B$ as in Refs.~\cite{Neupert2011,Yao2013,Wang2011}, and the next-next-nearest hopping $t_{3}$ is the same for both sublattices. In ${\kk}$-space, $a_{\kk}=\sum_{\rr}e^{i\kk\cdot\rr}a_{\rr}/\sqrt{N_s}$, $b_{\kk}=\sum_{\rr}e^{i\kk\cdot\rr}b_{\rr}/\sqrt{N_s}$, the Hamiltonian is given by
\begin{align}
  H_0=\sum_{\kk}\psi_{\kk}^{\dag}(h_{\kk}^{0}+h_{\kk}^{x}\sigma_x+h_{\kk}^{y}\sigma_y+h_{\kk}^{z}\sigma_z)\psi_{\kk},
\end{align}
where $\psi_{\kk}^{\dag}=(a_{\kk}^{\dag},b_{\kk}^{\dag})$, $h_{\kk}^{0}=-2t_3[\cos(k_x+k_y)+\cos(k_x-k_y)]$, $h_{\kk}^{z}=M+2t_2(\cos k_x-\cos k_y)$, $h_{\kk}^{x}=4t_1\cos\phi\cos\frac{k_x}{2}\cos\frac{k_y}{2}$, $h_{\kk}^{y}=4t_1\sin\phi\sin\frac{k_x}{2}\sin\frac{k_y}{2}$. The lowest Chern band energy is given by $E_{\kk}=h_{\kk}^{0}-\varepsilon_{\kk}$ with $\varepsilon_{\kk}=\sqrt{(h_{\kk}^{x})^{2}+(h_{\kk}^{y})^{2}+(h_{\kk}^{z})^{2}}$,
and the eigenstate given by
\begin{align}
  \chi_{\kk}=\begin{pmatrix}
-e^{-i\phi_{\kk}}\sin\frac{\theta_{\kk}}{2}\\
\cos\frac{\theta_{\kk}}{2},\\
\end{pmatrix}
\end{align}
where $\theta_{\kk}=\arccos(h_{\kk}^{z}/\varepsilon_{\kk}),\phi_{\kk}=\arctan(h_{\kk}^{y}/h_{\kk}^{x})$. The lowest Chern band becomes nearly flat under typical parameters such as $M=0,t_{2}=0.3t_{1},t_{3}=-0.2t_{1},\phi=\pi/4$, with a small band width $W\simeq0.1t_{1}$ and a large band gap $\Sigma\simeq2.5t_{1}$ separated from the upper Chern band.  It is characterized by the Chern number given by $\nu=\int d^{2}\kk\nabla\times A(\kk)/(2\pi)$, where the Berry's connection is given by $A(\kk)=i\chi_{\kk}^{\dag}\nabla_{\kk}\chi_{\kk}=(1-\cos\theta_{\kk})\nabla_{\kk}\phi_{\kk}/2$ with singularities at $\kk_{+}=(0,\pi)$ and $\kk_{-}=(\pi,0)$. Using Cauchy integral, we have
\begin{align}
  \nu=\oint A(\kk)\cdot\frac{d\kk}{2\pi}=-\frac{1}{4\pi}(\nu_{+}\cos\theta_{\kk_{+}}+\nu_{-}\cos\theta_{\kk_-})
\end{align}
where $\nu_{\pm}=\oint_{\pm}\nabla_{\kk}\phi_{\kk}\cdot d\kk=\pm2\pi$.  If $-4t_2<M<4t_{2}$, then $\cos\theta_{\kk_+}=-1$, $\cos\theta_{\kk_-}=+1$, the system is a Chern insulator with $\nu=1$; if $|M|>4t_{2}$, then $\nu=0$,  it is a trivial band insulator.
\section{Fractional Topological Phases}\label{frachall}
We study the case that all the dipole moments are aligned in z-direction by a strong external field. The dipole-dipole interaction is given by $V(\rr-\rr')=d^{2}/|\rr-\rr'|^{3}$ where $d$ is the dipole moment.  The dipolar interaction between nearest neighbors is the strongest given by $J=2\sqrt{2}d^{2}/(t_{1}a^{3})$. The model Hamiltonian,
\begin{align}
  H=H_0+\frac{1}{2}\sum_{\rr\neq\rr'}V(\rr-\rr')n_{\rr}n_{\rr'}.
\end{align}
To avoid self-interaction, we truncate the dipolar terms by considering the nearest distance between two sites on the periodic lattice $\psi(\rr+N_{x,y}a)=\psi(\rr)$, i.~e. the truncated order is $Jt_1/64$ on lattice with $N_x=N_y=4$. Thus the tail of dipolar interaction truncated by lattice size is very weak for $J\lesssim1$, due to its rapidly decaying. We explore the many-body ground state of $H$ by exactly diagonalizing a finite $N$-particle system at filling $\nu=N/N_s$ with the same lattice parameters $t_2=0.3t_1,t_3=-0.2t_1$ as mentioned in Sec.~\ref{chern}. Due to lattice translational symmetry, the total wavevector of the many-body state is conserved in the first Brillouin zone and we can classify the ground states in different total wavevector sectors. For convenience, we denote two dimensional wavevector $\kk=(k_{x},k_{y})$ in units of $2\pi/N_{x}a,2\pi/N_{y}a$.

\subsection{Degenerate Ground State Manifolds}
\begin{figure}
  \includegraphics[height=2.7in,width=3.4in]{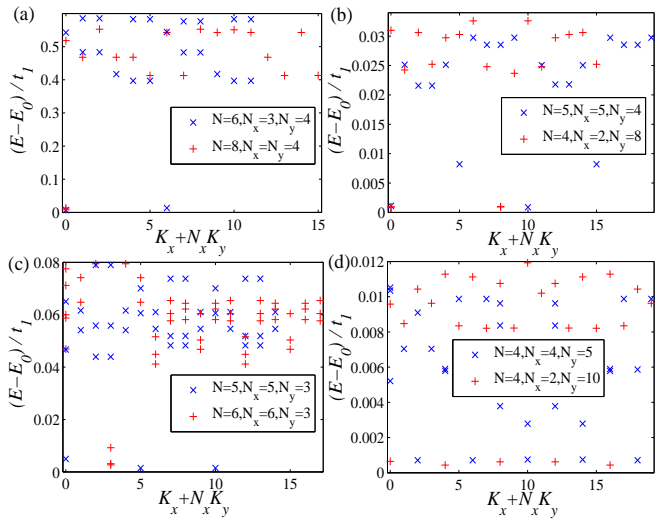}
  \caption{\label{energy}(Color online) Low-energy spectrum for bosonic $1/2,1/4$ and fermionic $1/3,1/5$ FQH states. (a)  $N=8,6$ bosons with $N_{x}=N/2,N_{y}=4$ at $J=0.5$. (b) $N=5$ bosons with $N_{x}=5,N_{y}=4$ and $N=4$ particles with $N_{x}=2,N_{y}=8$ at $J=0.5$. (c) $N=6,5$ fermions with $N_{x}=N,N_{y}=3$ at $J=0.5$. (d)  $N=4$ fermions with $N_{x}=4,N_{y}=5$ and $N=4$ fermions with $N_{x}=2,N_{y}=10$ at $J=0.6$. The energies are measured relative to the lowest energy for each system.}
\end{figure}

As shown in Fig.~\ref{energy}, these quasi-degenerate ground states at total wavector $K_i=(K_{ix},K_{iy})$ and filling $\nu=1/m$, can be qualitatively understood from the generalized Pauli principle~\cite{Regnault2011,Bergholtz2008}, i.~e. no more than one particle is allowed occupy within any consecutive $m$-orbitals with one dimensional orbital index $\lambda=k_x+N_xk_y$. For $N_{y}$ in multiples of $m$, this yields $K_{ix}=N(N_{x}-1)/2(\mathrm{mod} N_{x})$ and $K_{iy}=[N(i-1)+N(N_{y}-m)/2](\mathrm{mod} N_{y})$. For instance, two nearly degenerate states with total wavector $K_{1}=(0,0)$ and $K_{2}=(0,2$) for $N=6$ bosons on lattice with $N_{x}=3$, $N_{y}=4$ at $J=0.5$, separated from the higher energy levels by a large gap $\Delta\simeq0.4t_{1}$. The degeneracy of the ground states is equal to the inverse of the filling factor. In Sec. \ref{Flux Insertion}, we are going to show that these degenerate ground states are FQH states.

We characterize the robustness of the FQH ground states by the ratio of the ground energy splitting $\delta=E_{max}-E_{min}$ to the spectrum gap $\Delta=E_{lst}-E_{max}$, where $E_{max}(E_{min})$ is the maximal (minimal) energy in the degenerate manifold and $E_{lst}$ the lowest excited energy. The FQH states are more robust under a larger dipolar interaction with a smaller ratio of ground state energy splitting to the spectrum gap. These FQH states are rather robust for large $J$ but may not survive for small $J$ where the spectrum gap collapses. The only exception is $\nu=1/2$ boson case due to infinite onsite repulsion.  At higher fillings, for example at $\nu=2/3$, for $N=10$ dipolar fermions in a checkerboard lattice with $N_{x}=5$, $N_{y}=3$ at $J=0.5$, we do not find three robust and nearly degenerate states. The cases with lower fillings will be discussed in the last section.

\begin{figure}
  \includegraphics[height=2.8in,width=3.4in]{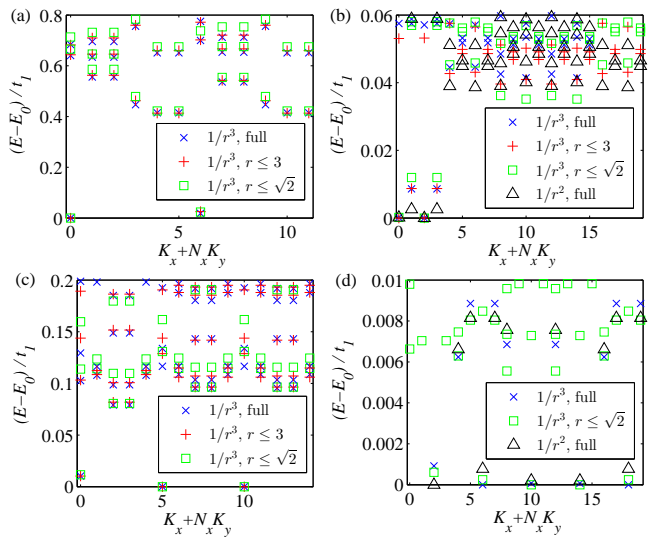}
  \caption{\label{energycompare}(Color online) Low-energy spectrum for bosonic $1/2,1/4$ and fermionic $1/3,1/5$ FQH states under long range $1/r^n$-power law decaying interaction truncated by different distances at the same ratio of nearest neighbor interaction to nearest hopping $J=1$: (a) $N=6$ bosons with $N_{x}=3,N_{y}=4$, (b) $N=5$ bosons with $N_{x}=4,N_{y}=5$, (c) $N=5$ fermions with $N_{x}=5,N_{y}=3$, (d) $N=4$ fermions with $N_{x}=4,N_{y}=5$. The energies are measured relative to the lowest energy for each system.}
\end{figure}

\begin{figure}
  \includegraphics[height=1.7in,width=3.4in]{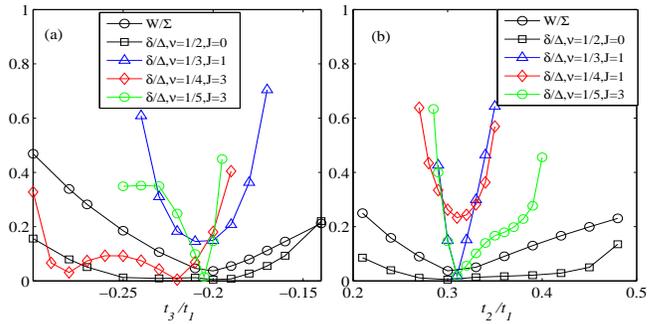}
  \caption{\label{bandwidth}(Color online) Robustness of $1/2,1/3,1/4$-FQH states via the ratio $\delta/\Delta$ as a function of the reduced flatness ratio $W/\Sigma$: (a) $t_3$ is changed while $t_2=0.3t_1$ is fixed to keep the single particle Berry curvature of lower Chern band unchanged; (b) $t_2$ is changed while $t_3=-0.2t_1$ is fixed. Fillings: $N=6$ bosons on lattice with $N_x=3,N_y=4$; $N=5$ fermions on lattice with $N_x=5,N_y=3$; $N=5$ bosons on lattice with $N_x=4,N_y=5$; $N=4$ fermions on lattice with $N_x=4,N_y=5$. }
\end{figure}

To elucidate the effects of long range interaction, we make a quantitative comparison with numerical results for the Hamiltonian where the dipolar interaction $1/r^3$ is truncated by a specific distance. As shown in Fig.~\ref{energycompare}, for various filling, the global picture of low energy spectrum is not strongly affected. However we find a slight decrease of the reduced ratio $\delta/\Delta$ as the interaction is truncated more earlier, demonstrating the enhancing stability of the FQH states in the presence of long range dipolar interaction on finite system. Also, the spectrum gap exceeds that obtained by Sheng's group~\cite{Sheng2011,Wang2011} with only nearest-neighbor and next-nearest-neighbor repulsion for these FQH states at the same interaction scale.  To further clarify the role played by the power-law tail of the long-range interaction, we consider a more slowly decaying interaction potential proportional to $1/r^n (2\leq n<3)$ for $\nu=1/4,1/5$-FQH states. Compared with those of $1/r^3$ potential, we get an even larger gap $\Delta$, quite sizable in Fig.~\ref{energycompare}(b) and (d), signifying the importance of long-range power-law stabilization of the FQH states.

The existence of FQH state is also affected by the band parameters, i.~e. the next-nearest hopping $t_{2}$ determining the topology of band structure and the next-next-nearest hopping $t_3$ contributing crucial to the flatness of the lowest Chern band. The characteristic measure for the flatness of Chern band is here the reduced flatness ratio $W/\Sigma$. FQH states do not always exist as ground states when tuning $t_2$ and $t_3$.  As shown in Fig.~\ref{bandwidth}, when $t_2$ and $t_3$ are tuned away from flatband region, the reduced flatness ratio $W/\Sigma$ increases and the ratio $\delta/\Delta$ becomes much larger, and eventually the degenerate ground state manifold disappears and merges into the dispersive bands. The optimal minimal point of $\delta/\Delta$ locates near the point $(t_2,t_3)=(0.3t_1,-0.2t_1)$, with deviation of the order of $0.01t_1$. We confirm that the FQH states are more robust when the system is in the flatband region. Our results are consistent with Wang {\it el al.}~\cite{Wang2011} and Kourtis {\it el al.}~\cite{Kourtis2012}, but do not compare quite well with that obtained by Grushin {\it et al.}~\cite{Grushin2012} who have drawn the conclusion from exact diagonalization of the projected Hamiltonian on lower Chern band. The nature of this discrepancy maybe stem from the fact that in Ref. \cite{Grushin2012} the effect of mixing with upper Chern band by the interaction is ignored, which is only valid when the interaction scale is much smaller than the band gap.
\subsection{Phase transitions induced by dipolar interactions}

\begin{figure}
  \includegraphics[height=2.4in,width=3.4in]{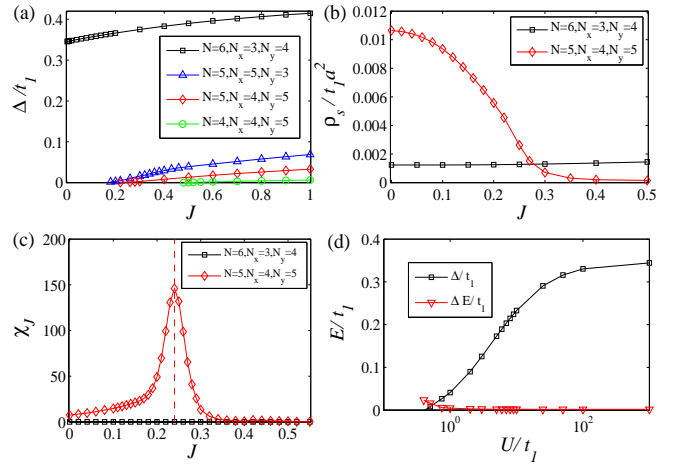}
  \caption{\label{energyJ}(Color online) (a) The spectrum gap $\Delta$ versus dipolar interaction strength $J$ for hardcore bosonic $1/2,1/4$ and fermionic $1/3,1/5$-FQH states. At fillings $\nu=1/3,1/4,1/5$, by decreasing the dipolar interaction, degenerate ground state manifolds eventually collapse.  At $\nu=1/2$, a quite large gap is present even without dipolar interaction, indicating the robustness of $1/2$-FQH state. (b) The superfluid density $\rho_s$ in the ground state at fillings $\nu=1/2,1/4$ as a function of $J$ for hardcore boson. In FQH states, $\rho_s$ vanishes up to a precision of $10^{-3}$. (c) The ground state wave function fidelity susceptibility for hardcore bosonic $1/2,1/4$-FQH states $\chi_J$ under the variation of $J$. The dashed vertical curves indicate quantum phase transition point. (d) DMRG results of the spectral gap $\Delta$ and ground energy splitting $\Delta E$ for softcore bosonic system with $N=6,N_x=3,N_y=4$ as a function of onsite repulsion $U/t_1$, without dipolar interaction.}
\end{figure}

As mentioned above, except for $\nu=1/2$ hardcore boson case, all FQH states that we consider vanish at some dipolar interaction strengths with collapse of the spectrum gap.  As shown in Fig.~\ref{energyJ}(a), we find that in the fermion case, the critical dipolar interaction strength is $J=0.18$ at filling $\nu=1/3$ and $J=0.47$ at filling $\nu=1/5$.  In the boson case, the critical dipolar interaction strength is $J=0.24$ at filling $\nu=1/4$.  Below these critical interactions, the fermion systems are in the Fermi liquid state and the boson systems are in the superfluid state.

In the boson case, the phase transition at the critical interaction strength is also indicated by the superfluid density.  In the superfluid state, there is a symmetry breaking in the order parameter $\langle a_{\rr}\rangle\propto \exp(i\phi(\rr))$, where $\phi(\rr)$ is the superfluid phase.  With twisted boundary condition $\psi(\rr+N_ya)=e^{i\theta}\psi(\rr)$ with very small $\theta$, the superfluid density $\rho_s$ can be determined from the energy relation $E(\theta)-E(0)\simeq\rho_{s}\theta^2/a^2$, where $E(\theta)$ is the ground state energy. Fig.~\ref{energyJ}(b) shows the evolution of $\rho_s$ in the lowest ground state as a function of $J$. Consistent with Fig.~\ref{energyJ}(a), at $\nu=1/2$, the superfluid density is zero and the system is a FQH state for any value of $J$. However, at $\nu=1/4$, a small but finite $\rho_s$ emerges when the dipolar interaction strength is reduced to $J=0.24$, signalling the quantum phase transition from the FQH phase into a superfluid phase.

We also calculate the ground-state fidelity susceptibility $\chi_J$, which measures the change of the ground state wave function $\psi(J)$ under a small change of the interaction strength, defined by
\begin{align}
  \chi_J=2\frac{1-F(J,\delta J)}{(\delta J)^2}
\end{align}
where the fidelity $F(J,\delta J)=|\langle\psi(J)|\psi(J+\delta J)\rangle|$ measures the overlap of the ground-state wavefunctions between $J$ and $J+\delta J$ with $\delta J\rightarrow0$. Inside a given quantum phase, the value of $\chi_J$ remains analytic and small. However, near a quantum phase transition point, the fidelity susceptibility diverges in the thermodynamic limit~\cite{Gu2010}, which serves as a signal of quantum phase transitions. As shown in Fig.~\ref{energyJ}(c), for bosonic $\nu=1/2$, there does not exist any peak in the fidelity susceptibility; for bosonic $\nu=1/4$, a peak near $J=0.24$ marks the phase transition between the Bose superfluid and the FQH state.

Finally, to clarify that $1/2$-FQH state originates from onsite repulsion only, we consider the Hubbard model for softcore Bosons in a finite system with $N=6,N_x=3,N_y=4$, and the model Hamiltonian is given by $H=H_0+\sum_{\rr}Un_{\rr}(n_{\rr}-1)/2$.  In the density matrix renormalization group (DMRG) approach, we keep up to $m=500$ basis states in DMRG block, test the performance by comparing three lowest energy states ($E_1,E_2,E_3$) with exact results in opposite limits $U=0$ and $U=\infty$, and obtain the accurate energies with energy deviations of the order $10^{-4}\sim10^{-3}t_1$ and the maximum truncation error less than $10^{-5}$. In Fig.~\ref{energyJ}(d), we show the evolutions of the energy spectral gap ($\Delta=E_3-E_2$) of two ground states $E_1,E_2$ separated from the third excited state $E_3$ and the energy splitting ($\Delta E=E_2-E_1$) of the gapped two ground states. For $U/t_1<(U/t_1)_c\simeq0.5$, the $1/2$-FQH state collapses due to the softening of $\Delta$ and the ground states on flatband are highly degenerate when $U=0$.
\subsection{Flux Insertion}\label{Flux Insertion}

\begin{figure}
  \includegraphics[height=2.7in,width=3.4in]{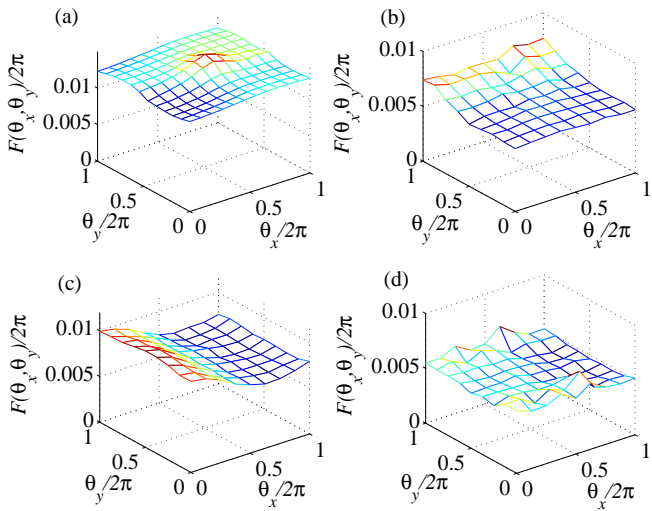}
  \caption{\label{berry}(Color online) Berry curvatures $F(\theta_x,\theta_y)/2\pi$ in: (a) the $K=(0,0)$ ground state of bosonic system with $N=6,N_x=3,N_y=4$; (b) the $K=(0,0)$ ground state of bosonic system with $N=5,N_x=4,N_y=5$; (c) the $K=(0,0)$ ground state of fermionic system with $N=5,N_x=5,N_y=3$; (d) the $K=(2,0)$ ground state fermionic system with $N=4,N_x=4,N_y=5$.}
\end{figure}

\begin{figure}
  \includegraphics[height=2.7in,width=3.4in]{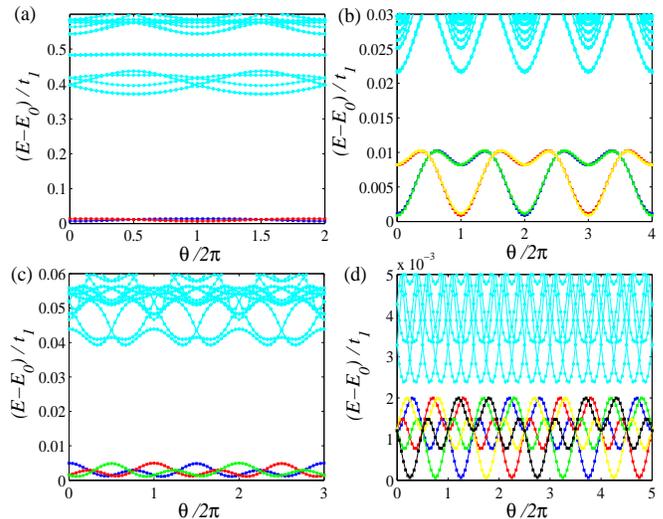}
  \caption{\label{energyflow}(Color online) The spectral flow under flux insertion along the y-direction, which is equivalent to the adiabatic insertion of flux quanta, for (a) $N=6$ bosons in a checkerboard lattice with $N_{x}=3,N_{y}=4$ at $J=0.5$, (b) $N=5$ bosons with $N_{x}=5,N_{y}=4$ at $J=0.5$, (c) $N=5$ fermions with $N_{x}=5,N_{y}=3$ at $J=0.5$, (d) $N=4$ fermions with $N_{x}=4,N_{y}=5$ at $J=0.6$.}
\end{figure}

The question whether or not these ground states are FQH states can be answered by their topological invariants, i.~e. the Chern numbers.  With twisted boundary conditions, $\psi(\rr+N_{x}a)=e^{i\theta_{x}}\psi(\rr)$,
$\psi(\rr+N_{y}a)=e^{i\theta_{y}}\psi(\rr)$ where $\theta_{x,y}$ are the twisted angle, the Chern number is given by $\nu=\int d\theta_{x}d\theta_{y}F(\theta_x,\theta_y)/2\pi$ where the Berry curvature is given by $F(\theta_x,\theta_y)=\mathbf{Im}(\langle{\partial_{\theta_x}\psi}|{\partial_{\theta_y}\psi}\rangle
-\langle{\partial_{\theta_y}\psi}{\partial_{\theta_x}\psi}\rangle)$. By calculating the Berry curvatures shown in Fig.~\ref{berry}, we obtain the Chern numbers of two gapped ground states for $\nu=1/2$ bosonic system with $N=6,N_x=3,N_y=4$ at $J=0.5$ and found $\nu_1\simeq0.51$ and $\nu_2\simeq0.49$; for fermionic system with $N=5,N_x=5,N_y=3$ at $J=0.5$, the Chern numbers of three gapped ground states at $\nu=1/3$ filling are $\nu_1\simeq0.32$, $\nu_2\simeq0.32$, and $\nu_3\simeq0.35$. The total Chern number of all the ground states of each system are unity, $\sum_{i=1}^{m}\nu_i=1$. Similarly, we obtain $\nu\simeq0.25$ for the $K=(0,0)$ ground state of bosonic system with $N=5,N_x=4,N_y=5$ and $\nu\simeq0.20$ for the $K=(2,0)$ ground state of fermionic system with $N=4,N_x=4,N_y=5$.  We find that the fractional Chern number holds true even in thin torus case, i.e. $\nu\simeq0.24$ for the $K=(0,0)$ ground state of bosonic system with $N=5,N_x=2,N_y=10$ and $\nu\simeq0.20$ for the $K=(0,0)$ ground state of fermionic system with $N=4,N_x=2,N_y=10$.  By expanding the many body wave function in the lowest Chern band $\psi=\sum_{\{\kk_{i}\}}\psi(\{\kk_{i}\})\prod_{i=1}^{N}\chi_{\kk_i}$, the Chern number can be further written as
\begin{align}
  \nu&=\frac{i}{2\pi}\int d^{2}\kk n_{\kk}\nabla_{\kk}\times(\chi_{\kk}\nabla_{\kk}\chi_{\kk})\nonumber\\
  &=-\frac{1}{4\pi}\left[n_{k_+}\cos\theta_{k_+}\nu_{+}+n_{k_-}\cos\theta_{k_-}\nu_{-}\right].
\end{align}
The momentum occupation number of the many body ground state in the lowest Chern band is given by $n_{\kk}=\langle\psi|\beta_{\kk}^{\dag}\beta_{\kk}|\psi\rangle=\sum_{\kk_{i}}'\psi^{2}(\{\kk_{i}\})$ where $\beta_{\kk}$ is the annihilation operator in the lowest Chern band and the summation is only over the configuration with ${\kk}$-state occupied. In the thermodynamic limit the orbital occupation should be uniform, $n_{k_+}=n_{k_-}=1/m$, consistent with the $\nu=1/m$ FQH state.

\begin{figure}
  \includegraphics[height=1.6in,width=3.4in]{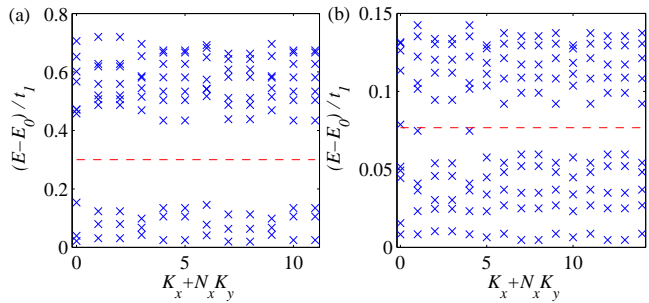}
  \caption{\label{quasi}(Color online) The low-energy quasihole spectrum. Only ten lowest energies per momentum sector are displayed. (a) $N=5$ bosons on lattice with $N_{x}=3,N_{y}=4$ at $J=0.5$. The number of states below the red dashed line is 3 per momentum sector. (b) $N=4$ fermions on lattice with $N_{x}=5,N_{y}=3$ at $J=0.5$. The number of states below the red dashed line is 5 per momentum sector.}
\end{figure}

In addition, topological degeneracy of the ground state manifold should be $m$-fold for $\nu=1/m$ filling. By varying $\theta_{y}$ from 0 to $2m\pi$ which is equivalent to the adiabatic insertion of $m$ flux quanta into the system~\cite{Neupert2011,Thouless1989}, these quasi-degenerate ground states evolve into each other without mixing with exited levels during the spectra flow.  As shown in Fig.~\ref{energyflow}, the ground states of our system clearly show the robustness of topological degeneracy. Adiabatically inserting $m$ flux quanta changes the Berry phase associated with braiding a quasiparticle around the circle by $\Delta\theta=2\pi m q$~\cite{Levin2009}, where $q$ is the charge of the quasiparticle. Since the final state is in the same topological sector with the initial state, $\Delta\theta$ must be an integer multiple of $2\pi$, thus the smallest charge of the quasiparticle should be $q=1/m$. One of the hallmarks of the FQH state is the existence of quasiholes excitations which carry fractional charge obeying the fractional statistics.  Removing one particle or adding $m$-flux quanta would create $m$ quasiholes, each with charge $1/m$.  From Ref.~\cite{Regnault2011}, the number of quasihole states of $N$ particles in $N_{s}$ orbitals reads as $N_q=N_s\times[N_s-(m-1)N-1]\text{!}/[N_s-mN]\text{!}/N\text{!}$. As shown in Fig.~\ref{quasi}, we compute the spectrum of quasiholes which lies in a low-energy manifold (quasihole states) separated by a gap from higher states, and find their number matches theoretical analysis.
\subsection{Density Structure Factor}

\begin{figure}
  \includegraphics[height=2.7in,width=3.4in]{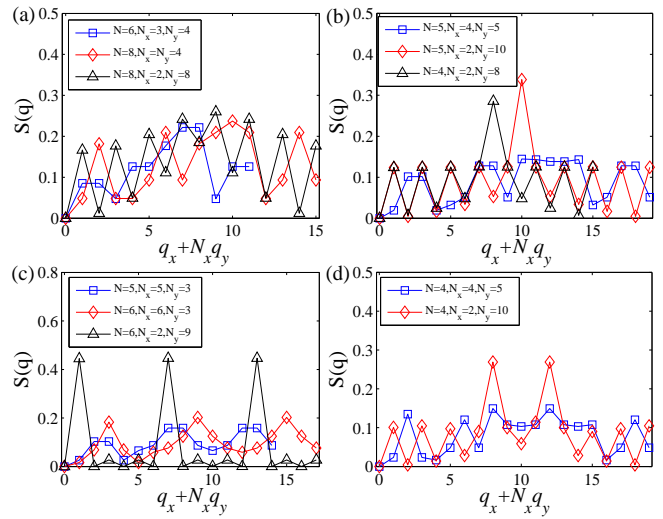}
  \caption{\label{structure}(Color online) The static density structure factors $S({\bf q})$ in: (a) $\nu=1/2$ bosonic states at $J=0.5$; (b) $\nu=1/4$ bosonic states at $J=0.5$; (c) $\nu=1/3$ fermionic states at $J=0.5$ ($J=200$ for $N=6,N_{x}=2,N_{y}=9$);  (d) $\nu=1/5$ fermionic states at $J=0.6$. The structure factors are almost the same for the ground states in the same degenerate manifold.  The Bragg peaks in structure factors are indications of CDW states.}
\end{figure}

Another important issue is whether or not there is a commensurate CDW state competing with FQH states~\cite{Grushin2012}.  The m-fold degeneracy of ground states alone does not resolve this issue, due to possible degenerate CDW states. One distinctive feature of the CDW state is the Bragg peak in the static density structure factor defined by
\begin{align}
  S(\qq)=\frac{1}{2N_{s}}\sum_{\rr,\rr'}e^{i\qq\cdot(\rr-\rr')}\left(\langle n_{\rr}n_{\rr'}\rangle-\langle n_{\rr}\rangle\langle n_{\rr'}\rangle\delta_{\qq,0}\right).
\end{align}
As shown in Fig.~\ref{structure}, in our numerical results of finite two-dimensional systems, there are no particular Bragg peaks in $S(\qq)$ at any finite wavevector.  We also calculate the local densities of these degenerate states and they are almost uniform for both sublattice, $\langle a_{\rr}^{\dag}a_{\rr}\rangle\simeq\langle b_{\rr}^{\dag}b_{\rr}\rangle\simeq\nu/2=N/2N_s$ with error less than two percent. The intrasublattice structure factor $S^{aa}(\qq)=\sum_{\rr,\rr'}e^{i\qq\cdot(\rr-\rr')}\left(\langle n_{\rr}^{a}n_{\rr'}^{a}\rangle-\langle n_{\rr}^{a}\rangle\langle n_{\rr'}^{a}\rangle\delta_{\qq,0}\right)/N_{s}$ and the intersublattice structure factor $S^{ab}(\qq)=\sum_{\rr,\rr'}e^{i\qq\cdot(\rr-\rr')}\left(\langle n_{\rr}^{a}n_{\rr'}^{b}\rangle-\langle n_{\rr}^{a}\rangle\langle n_{\rr'}^{b}\rangle\delta_{\qq,0}\right)/N_{s}$ do not exhibit any peak as well. Thus we can rule out the possibility of CDW states as the competing ground state.

Nevertheless, in the one-dimensional thin torus limit, the ground states at rational filling fraction $\nu=1/q$ are not only q-fold degenerate, but also display peaks in $S(\qq)$.  We find that: (\textrm{i}) For $N=4,5$ bosons on lattice with $N_{x}=2,N_{y}=2N$ at $J=0.5$ exhibiting one Bragg peak at $\qq=(0,N)$; (\textrm{ii}) for $N=4$ fermions on lattice with $N_{x}=2,N_{y}=10$ at $J=0.6$ exhibiting two Bragg peaks at $\qq_1=(0,4)$ and $\qq_2=(0,6)$; (\textrm{iii}) For strongly interacting $N=6$ fermions on lattice with $N_{x}=2,N_{y}=9$ at $J=200$ exhibiting three Bragg peaks at $\qq_1=(0,4)$, $\qq_2=(0,6)$ and $\qq_3=(1,0)$, where the $\qq_3=(1,0)$ peak comes from $S^{aa}(\qq)$ and the other two come from both $S^{aa}(\qq)$ and $S^{ab}(\qq)$. These imply the one dimensional crystalline order with the small aspect ratio $N_x/N_y$ and the strong interaction strength, as demonstrated in the continuum thin-torus limit where the crystallized phase is adiabatically connected to 2D FQH states~\cite{Seidel2005}. These fascinating CDW states in the thin torus lattice survive even under dipolar interaction truncated by shorter distance and can be interpreted as a symmetry-protected topological phase~\cite{Bernevig2012,Grusdt2014}, due to the inversion symmetry respected by many-body FQH model wave functions~\cite{WRB2012}.
\section{Summary and Discussions}\label{SD}
In summary, we show that dipolar quantum gas could host FQH states at a partial filling $\nu=1/m$ in the lowest Chern band, characterized by $m$-fold quasidegenerate ground states, with very small energy splitting owing to the finite-size effects. Numerically we have found convincing evidences of the fermionic $1/3,1/5$ and the bosonic $1/2,1/4$-FQH states, protected by a robust energy gap from higher excited states. The spectrum gap can be enhanced by increasing dipolar interaction strength, such that FQH states may be desirable by trapping dipolar gas in optical lattices with Chern bands.  In realistic situations, inhomogeneities, such as impurities and disorder, are expected.  Weak disordered onsite potential could cause level anti-crossing among these ground states under the adiabatic insertion of flux quanta. However, the spectral gap between these nearly degenerate states and the first excited state should remain open.

We also find evidence for FQH states with lower fillings. Numerical studies on lattice with the same parameters and $J=2$, indicate that: (\textrm{i}) six quasidegenerate gapped states with total wavevector $K_i=(0,i),i=0-5$ separated from the higher energy levels by a gap $\Delta\simeq0.0074t_{1}$, emerge for $N=5$ dipolar bosons on lattice with $N_x=5,N_y=6$; (\textrm{ii}) seven quasidegenerate gapped states with total wavevector $K_i=(0,i),i=0-6$ separated from the higher energy levels by a gap $\Delta\simeq0.0055t_{1}$, emerge for $N=5$ dipolar fermions on lattice with $N_x=5,N_y=7$.  These results indicates that the checkerboard model should host robust bosonic $1/6$ and fermionic $1/7$-FQH states for stronger dipolar interaction strengths. With three body interactions polar molecules~\cite{BMZ2007} in this type of optical lattice may be in a fermionic non-Abelian FQH states similar to Moore-Read phases at $\nu=\frac{1}{2}$ filling~\cite{WBR2012,LB2013}.

There are several promising ways to realize fractional Chern states in experiments.  In the proposal for $^{40}\text{K}^{87}\text{Rb}$ system by Yao {\it et al.}~\cite{Yao2013}, local spatial optical dressing couples the three lowest excited rovibrational states, spanning a pseudo-spin Hilbert space with the rovibrational ground state. The projected dipolar spin-exchange interaction between two sites $V_{dd}=[J_{ij}^{z}S_{i}^{z}S_{j}^{z}+\frac{1}{2}(J_{ij}^{\perp}S_{i}^{+}S_{j}^{-}+h.c.)]/|r_i-r_j|^3$. Thus, when described by hard-core boson operators, the effective hopping of reduced bosonic checkerboard model $t_{\rr,\rr'}\simeq[d_{00}^{2}w_{\rr}^{\ast}w_{\rr'}-\frac{1}{2}d_{01}^{2}(v_{\rr}^{\ast}v_{\rr'}+s_{\rr}^{\ast}s_{\rr'})]/|\rr-\rr'|^3$ where the dipolar interaction strength $d_{00}^{2}/a^3\sim d_{01}^{2}/a^3\sim400$hHz in a 340nm-lattice~\cite{Yan2013} and optical dressing parameters $w,v,s$ are tunable. The kinetic hopping and residual dipolar interaction are of the same energy scale.  A flatband with $t_1\sim20$nK and a spectral gap of $\Delta\sim2$nK for 1/2-FQH states are expected when one takes the same parameters as Ref.~\cite{Yao2013}.  By tuning the tilted angle of electric field and $s\ll v,w$, the dipolar interaction exhibit strong anisotropy and the effective bosonic hopping of band structure is changed, inducing a much richer phase diagram including a superfluid, a CDW and a 1/2-FQH state~\cite{Yao2013,Barkeshli2014}.  Another desirable approach to realize fractional Chern insulator is to trap ultracold dipolar gases such as $^{40}\text{K}^{87}\text{Rb}$ and $^{161}\text{Dy}$, in the Hofstadter-like optical lattice~\cite{Aidelsburger2014}, due to the adiabatic continuity between Hofstadter and Chern insulator states~\cite{Wu2012}. When the spin degree is considered (e.~g., the magnetic dipolar interaction among different components of $\text{Dy}$), one would expect many more exotic possible phases to occur, like flatband ferromagnetism, fractional quantum spin Hall states~\cite{LSTC2014} and so on.

\textit{Note added.} At the time of preparing the second version of this manuscript, we became aware of a preprint on the effects of dipolar repulsion interaction on non-Abelian FQH states mediated by nearest neighbor attraction~\cite{Wang2014}.

\begin{acknowledgments}
T.S.Z wishes to thank Hong Yao for useful discussions. This work is supported by NSFC under Grant No 11274022.
\end{acknowledgments}

\end{document}